\documentclass[
11pt,tightenlines,%
nofootinbib,preprintnumbers,prd]{revtex4}
\usepackage{graphicx}
\newcommand{\be}{\begin{equation}}
\newcommand{\ee}{\end{equation}}
\newcommand{\bea}{\begin{eqnarray}}
\newcommand{\eea}{\end{eqnarray}}
\newcommand{\nn}{\nonumber}
\newcommand{\0}{\over }
\newcommand{\2}{{1\over2}}

\begin{document}
\preprint{TUW-02-20}
\title{Renormalization Group Summation and the Free Energy
of Hot QCD}
\author{D. G. C. McKeon}
\affiliation{Institut f\"ur Theoretische Physik, Technische
Universit\"at Wien, 
A-1040 Vienna, Austria }
\affiliation{Dept. of Applied Mathematics, University
of Western Ontario, London, Ontario N6A 5B7, Canada}
\author{A. Rebhan}
\affiliation{Institut f\"ur Theoretische Physik, Technische
Universit\"at Wien, 
A-1040 Vienna, Austria }
\begin{abstract}
Using an approach developed in the context of zero-temperature QCD
to systematically sum higher order effects whose form is fixed
by the renormalization group equation, we sum to all orders
the leading log (LL) and next-to-leading log (NLL) contributions
to the thermodynamic free energy in hot QCD.
While the result varies considerably less with changes in
the renormalization scale than does the purely perturbative result,
a novel ambiguity arises
which reflects the strong scheme dependence of
thermal perturbation theory.
\end{abstract}
\maketitle

\section{Introduction and Summary}

The renormalization procedure in quantum field theory inevitably
introduces a renormalization scale parameter $\mu^2$ into perturbative results;
as $\mu^2$ is unphysical, its value is in principle arbitrary.
Consequently, the calculation of any physical quantity would necessarily
result in the dependence on $\mu^2$ disappearing.
However, at any finite order of perturbation theory, residual
dependence on $\mu^2$ renders the result ambiguous, for changing
the value of $\mu^2$ changes the predicted value of the physical
quantity that has been computed \cite{Stevenson:1981vj}.

This problem has proved to be particularly acute in the calculation
of the thermodynamic free energy in thermal field theory \cite{Kap:FTFT}.
In quantum chromodynamics (QCD) at temperatures much larger than
the deconfinement temperature the free energy has been calculated
\cite{Arnold:1995eb,Zhai:1995ac,Braaten:1996jr}
to order $\alpha_s^{5/2}$, four terms in the perturbative series
beyond the leading ideal-gas term. Whereas the first few approximations
turn out to show little sign of convergence for any temperature
of practical interest, the result to order $\alpha_s^{5/2}$
happens to be centered about the results obtained in lattice
gauge theory, but the dependence on the renormalization
scale parameter $\mu^2$ is so large that it has little
predictive power.

Recently, in Ref.~\cite{Ahmady:2002fd} in the context of
standard model calculations, it was shown how the renormalization
group (RG) equation can be used to sum in a systematic manner
the leading log (LL), next-to-leading log (NLL), \ldots effects given
the perturbative results to one loop, two loop, \ldots order.
This so-called ``renormalization group summation'' (RG$\Sigma$)
has been found to lead to a considerable reduction of the
dependence on the parameter $\mu^2$ within a given renormalization scheme.

In the case of thermal field theory this procedure requires
generalization because the perturbative series also involves
half-integer powers and logarithms of $\alpha_s$. This is
carried out in Sec.~\ref{sec:RGS} and applied to the
available three-loop result for the free energy in hot QCD
in the modified minimal subtraction ($\overline{{\rm MS}}$) scheme.
The RG$\Sigma$ result which includes all LL and NLL effects turns out
to be only weakly dependent on $\mu^2$. Unfortunately, as
discussed in Sec.~\ref{sec:disc}, this does
not increase substantially the predictive power of the three-loop calculation,
because even within a fixed renormalization scheme (here the
$\overline{{\rm MS}}$ scheme) there arises a new ambiguity
in the form of the initial conditions for the differential
equations, whose integration carries out the RG-summation.
This ambiguity is somewhat larger than at zero temperature
because the RG-summation of the thermal perturbative series
leads to two uncoupled sets of differential equations rather
than one, reflecting the particular difficulties that
thermal perturbation theory present.

In the Appendix, we briefly discuss the difference between
the strictly perturbative solution for the running coupling $\alpha_s$
to two-loop order and the exact solution at two-loop order,
which can be given in closed form in terms of Lambert's W function.

\section{Renormalization Group Summation in Hot QCD}
\label{sec:RGS}

In QCD with $n_f$ flavors of quarks, the thermodynamic free energy
at high temperature has been computed to be 
\cite{Arnold:1995eb,Zhai:1995ac,Braaten:1996jr}
\bea\label{Fpt}
{\cal F} &=& {-8\pi^2\045}T^4 \biggl\{ 
\left(1+{21\032}n_f\right)+
{-15\04}\left(1+{5\012}n_f\right){\alpha_s\0\pi}
+30\left[\left(1+{n_f\06}\right)\left({\alpha_s\0\pi}\right)\right]^{3/2}\nn\\
&&+\Bigl\{237.2+15.97n_f-0.413n_f^2+
{135\02}\left(1+{n_f\06}\right)\ln\left[{\alpha_s\0\pi}(1+{n_f\06})\right]\nn\\
&&\qquad\qquad-{165\08}\left(1+{5\012}n_f\right)\left(1-{2\033}n_f\right)
\ln{\bar\mu\02\pi T}
\Bigr\}\left({\alpha_s\0\pi}\right)^2\nn\\
&&+\left(1+{n_f\06}\right)^{1/2}\biggl[-799.2-21.96 n_f - 1.926 n_f^2\nn\\
&&\qquad\qquad+{495\02}\left(1+{n_f\06}\right)\left(1-{2\033}n_f\right)
\ln{\bar\mu\02\pi T}\biggr]\left({\alpha_s\0\pi}\right)^{5/2}+
\mathcal O(\alpha_s^3\ln\alpha_s) \biggr\}
\eea
where $\bar\mu$ is the renormalization scale parameter of
the $\overline{{\rm MS}}$ scheme and $\alpha_s(\bar\mu)$ is the
running coupling in this scheme whose form to three-loop order
is given in \cite{Groom:2000in}, though we shall restrict ourselves
to its two-loop version in the following (see the Appendix for
more discussion).

Changing the renormalization scale parameter $\bar\mu$ in principle
should not alter the value of $F$, as changes in the explicit $\bar\mu$
are compensated for by changes of $\alpha_s(\bar\mu)$,
and indeed, the result (\ref{Fpt})
is independent of $\bar\mu$ to order $\alpha_s^{5/2}$. However,
numerically the dependence on $\bar\mu$ is large, in fact larger
than that of the result to order $\alpha_s^1$ unless $\bar\mu$,
which has to be of the same order as $T$ to avoid large logarithms,
is much larger than even the electroweak scale.

Extrapolating from Eq.~(\ref{Fpt}), we assume the complete all-order
result of $\cal F$ can be represented by a series of the form
\be\label{FRST}
\mathcal F/\mathcal F_0=1+\sum_{n=0}^\infty
\left( R_n(u)x^{n+1}+S_n(u) x^{n+{3\over2}}+T_n(u)x^{n+2}\ln x \right)
\ee
where $\mathcal F_0$ is the ideal-gas value,
$x=\alpha_s(\bar\mu)/\pi$, $u=xL$, $L=\ln(\bar\mu^2/(2\pi T)^2)$
and
\be\label{RST}
R_n(u)=\sum_{m=0}^\infty A_{n+m,m}u^m,\quad
S_n(u)=\sum_{m=0}^\infty B_{n+m,m}u^m,\quad
T_n(u)=\sum_{m=0}^\infty C_{n+m,m}u^m,
\ee
although only the coefficients with $n\le 1$ are accessible
by thermal perturbation theory. In fact, all of the
perturbatively accessible cofficients with $m=0$ have been calculated
already, with the exception of $C_{1,0}$, which
is forthcoming\footnote{Y. Schr\"oder, private communication}.

In Ref.~\cite{Ahmady:2002fd} it has been shown how to sum all
RG-accessible logarithms when the lowest-order coefficients
to the sums in Eq.~(\ref{RST}), $A_{n,0}$, $B_{n,0}$, and
$C_{n,0}$, and the $\beta$-function coefficients
(see Eq.~(\ref{beta}) below) are known.
In this paper we extend this to a perturbative series of the
form (\ref{FRST}).

While the perturbative expression (\ref{Fpt}) is in powers of
$x^{1/2}$, successive RG$\Sigma$-perturbative
expressions are given by
\bea
\mathcal F^{(1)}_{\rm RG\Sigma}/\mathcal F_0&=&1+x R_0(xL),\\
\mathcal F^{(2)}_{\rm RG\Sigma}/\mathcal F_0&=&1+x R_0(xL)+x^{3/2}S_0(xL),\\
\mathcal F^{(3)}_{\rm RG\Sigma}/\mathcal F_0&=&1+x R_0(xL)+x^{3/2}S_0(xL)
+x^2(R_1(xL)+T_0(xL)\ln x),\\
\mathcal F^{(4)}_{\rm RG\Sigma}/\mathcal F_0&=&1+x R_0(xL)+x^{3/2}S_0(xL)
+x^2(R_1(xL)+T_0(xL)\ln x)+x^{5/2}S_1(xL)
\label{N3LRGS}\eea
which all are perturbatively accessible in hot QCD, as we shall see.

The explicit dependence of $\mathcal F$ on $\bar\mu^2$ and its implicit
dependence through $x(\bar\mu^2)$ are such that
\be\label{dF0}
\bar\mu^2 {d\mathcal F\over d\bar\mu^2}=0=\left(\bar\mu^2 {\partial \over
\partial \bar\mu^2 } + \beta(x) {\partial \over
\partial x} \right) \mathcal F
\ee
where
\be\label{beta}
\beta(x)=\bar\mu^2 {\partial x \over
\partial \bar\mu^2 }=(b_2 x^2+b_3 x^3+b_4 x^4 + \ldots ).
\ee
Substitution of (\ref{FRST}) into (\ref{dF0}) yields
\bea
0&=&\sum_{n=0}^\infty \biggl\{ \left[ R_n' x^{n+2}
+(b_2 x^2 + \ldots )(uR_n'+(n+1) R_n + x T_n) x^n \right] \nn\\
&&+\left[ S_n' x^{n+{5\02}}
+(b_2 x^2 + \ldots )(uS_n'+[n+{3\02}] S_n) x^{n+\2} \right] \nn\\
&&+\left[ T_n' x^{n+3}
+(b_2 x^2 + \ldots )(uT_n'+(n+2) T_n) x^{n+1} \right]\ln x \biggr\}
\eea

Lowest order terms of the form $x^n$, $x^{n+\2}$ and $x^n \ln x$ give rise
to differential equations for $R_0(u)$, $S_0(u)$ and $T_0(u)$, respectively.
The boundary conditions on these equations are the computed values
of $A_{0,0}$, $B_{0,0}$ and $C_{0,0}$ respectively, which can be
read off Eq.~(\ref{Fpt}). This gives
\be
R_0(u)=A_{0,0}w^{-1},\;
S_0(u)=B_{0,0}w^{-3/2},\;
T_0(u)=C_{0,0}w^{-2},\quad w\equiv (1+b_2 u).
\ee
These functions incorporate the LL contributions to $\mathcal F$ to
all orders.

To next order in the coupling $x$, we find differential equations for
$R_1$, $S_1$, and $T_1$ (which rely on knowning the above solutions
for $R_0$, $S_0$, and $T_0$). Solving these equations gives
\bea
R_1&=& w^{-2} \left[ A_{1,0}-\left({b_3\over b_2}A_{0,0}+C_{0,0} \right)
\ln w \right]\\
S_1&=&w^{-5/2} \left[ B_{1,0}-{3\over2}{b_3\over b_2}B_{0,0} \ln w \right]\\
T_1&=&w^{-3} \left[ C_{1,0}-2{b_3\over b_2}C_{0,0} \ln w \right]
\eea
which incorporate the NLL contributions to $\mathcal F$ to all orders.

Continuing in this way, we obtain
\bea
R_2&=&w^{-3}\biggl[ A_{2,0}-
\left(2{b_3\0b_2}A_{1,0}+{b_3^2\0b_2^2}A_{0,0}+C_{1,0}+{b_3\0b_2}C_{0,0}
\right)\ln w \nn\\
&&\qquad\quad + {b_3\0b_2}\left({b_3\0b_2}A_{0,0}+2C_{0,0}\right)\ln^2 w
+\left({b_3^2\0b_2}-b_4\right)A_{0,0}\,u \biggr]
\\
S_2&=&w^{-7/2}\biggl[ B_{2,0}-{b_3\0b_2}\left({5\02}B_{1,0}+
{3\02}{b_3\0b_2}B_{0,0} \right)\ln w 
+{15\08}{b_3^2\0b_2^2}B_{0,0}\ln^2 w+
{3\02}\left({b_3^2\0b_2}-b_4\right)B_{0,0}\,u \biggr]
\\
T_2&=&w^{-4}\biggl[ C_{2,0}-{b_3\0b_2}\left(       3C_{1,0}+
2{b_3\0b_2}C_{0,0} \right)\ln w 
+        3{b_3^2\0b_2^2}C_{0,0}\ln^2 w+
       2\left({b_3^2\0b_2}-b_4\right)C_{0,0}\,u \biggr]
\eea
these give the NNLL contributions to $\mathcal F$.

The coefficients $A_{0,0}$, $B_{0,0}$, $C_{0,0}$, $A_{1,0}$, and
$B_{1,0}$ are determined in the $\overline{{\rm MS}}$ scheme
by Eq. (\ref{Fpt}) and this allows us to
construct the RG$\Sigma$ approximation $\mathcal F^{(4)}_{\rm RG\Sigma}$ given
in Eq.~(\ref{N3LRGS}). The next approximation would also involve
$T_1$, $R_2$, $S_2$, and $T_2$. For these, one would
need results for $C_{1,0}$ and $A_{2,0}$, $B_{2,0}$, $C_{2,0}$,
but only $C_{1,0}$ is computable in perturbation theory
because the order $\alpha_s^3$-contribution to $\mathcal F$
is inherently nonperturbative \cite{Linde:1980ts,Gross:1981br}.
The remaining coefficients would have to be derived from
a nonperturbative framework such as lattice gauge theory, or
they might be estimated by nonperturbative resummation techniques
such as Pad\'e approximations \cite{Kastening:1997rg,Hatsuda:1997wf}.
Given that, the above formulae would allow the construction of
the next RG$\Sigma$ approximation $\mathcal F^{(5)}_{\rm RG\Sigma}$.

\section{Discussion}
\label{sec:disc}

\begin{figure}
\includegraphics[bb=50 205 540 545,width=10cm]{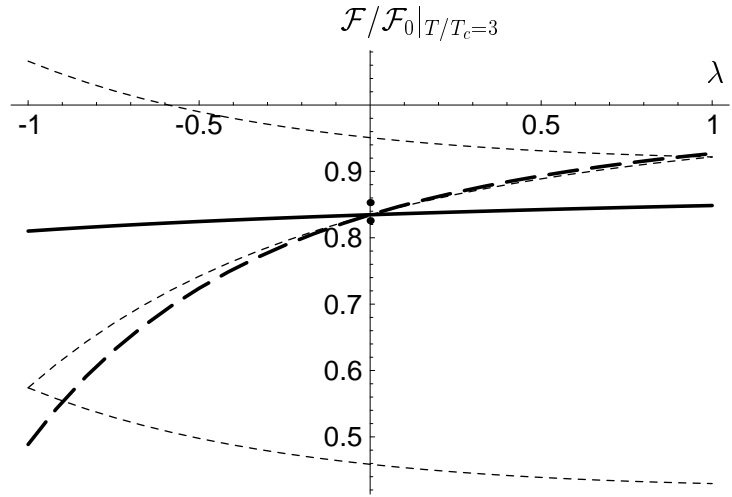}
\caption{\label{figkappab} Comparison of the renormalization-scale
dependence of the RG$\Sigma$ result 
for $\mathcal F/\mathcal F_0$ at $T=3T_c$ for $n_f=0$ 
(solid line) and the
perturbative result to order $\alpha_s^{5/2}$ (long-dashed line)
when varying
the renormalization scale $\bar\mu$ around a central value of
$2\pi T$ by a factor $e^\lambda$. The two dots on the vertical axis give two recent lattice results
from Refs.~\cite{Boyd:1996bx,Okamoto:1999hi}.
The short-dashed lines forming a big Z show the dependence of the
RG$\Sigma$ result on varying the arbitrary parameters $\kappa$ and
$\nu$ by a factor of $e^\lambda$ around 1.
}
\end{figure}

In Fig.~\ref{figkappab} the renormalization scale dependence
of the purely perturbative result (\ref{Fpt}) 
for $\mathcal F/\mathcal F_0$ at $T=3T_c$ for $n_f=0$ 
is displayed using
a two-loop running coupling constant $\alpha_s(\bar\mu=e^\lambda 2\pi T)$
with $\lambda$ varying between $-1$ and 1 (long-dashed line).
The two dots on the vertical axis give two recent lattice results
from Refs.~\cite{Boyd:1996bx,Okamoto:1999hi},
and the perturbative result is seen to agree well with these
for $\bar\mu=2\pi T$ (i.e. $\lambda=0$), but deviates strongly
for different choices of the renormalization scale.

The RG$\Sigma$ result as obtained above is given by the solid line,
and it shows a rather weak dependence on the renormalization scale.
The systematic RG summation as described above is thus able
to absorb almost all of the scale dependence.

However, this does not mean that the RG$\Sigma$ approach
predicts the perturbative result at $\bar\mu=2\pi T$. Rather,
the latter has been used as the intial condition for the
differential equations determining the functions $R_0$, $S_{0}$,
$T_0$, $R_1$, and $S_1$. Organizing the RG summation in terms
of a variable $L_\kappa=\ln[\bar\mu^2/(\kappa 2\pi T)^2]$
instead of the variable $L$ introduced in Eq. (\ref{FRST})
would have led to a different definition of the
constants $A_{0,0}$, $B_{0,0}$, $C_{0,0}$ that provide
the initial conditions. Moreover, the series involving
half-integer powers of $x$ in Eq.~(\ref{FRST}) leads to
differential equations that decouple from those responsible
for the series involving integer powers and logarithms of $x$.
The former series could therefore have been derived by introducing
a different variable $L_\nu=\ln[\bar\mu^2/(\nu 2\pi T)^2]$
for $S_n$.

In identifying the ambiguity in $L_\kappa$ and $L_\nu$ (as
parametrized by $\kappa$ and $\nu$ respectively) we are keeping
the value of $x$ unaltered---both $\bar\mu$ is held fixed and the
form of $x(\bar\mu)$ dictated by having chosen to work in the
$\overline{{\rm MS}}$ scheme is not changed. Rather, we note
that the invariance of the perturbative result (\ref{Fpt})
under the change
\be
A_{1,0}+A_{1,1}\ln{\bar\mu\02\pi T} \to
(A_{1,0}+A_{1,1}\ln\kappa)+A_{1,1}\ln{\bar\mu\0\kappa 2\pi T}
\equiv A_{1,0}^{(\kappa)}+A_{1,1}L_\kappa
\ee
(with analogous equations for $B_{1,0}^{(\nu)}$ and $C_{1,0}^{(\kappa)}$)
is lost when one deals with the RG$\Sigma$ expressions $R_1$, $S_1$
and $T_1$. The boundary condition for the differential equations
for $R_1$, $S_1$ and $T_1$, and the logarithm in the solution to these
equations have a dependence on $\kappa$ (or $\nu$) that no longer
automatically compensates as in the perturbative result.

In Fig.~\ref{figkappab} the short-dashed lines forming a big Z
show the result of varying $\kappa$ and $\nu$ around the value
1 by a factor $e^\lambda$ with $\lambda$ between $-1$ and $1$.
The upper bar in the big Z is formed by $\nu=e^{1}$,
$\kappa=e^\lambda$, the diagonal by $\nu=\kappa=e^\lambda$, and
the lower bar by $\nu=e^{-1}$, $\kappa=e^\lambda$.
Evidently, when $\kappa$ and $\nu$ are identified, the
ambiguity is a little bit smaller than the one given by the
renormalization scale dependence of the purely perturbative
result, but varying $\nu$ independently of $\kappa$ leads
to even larger variations, at their extremes even exceeding
the ideal gas result for $|\mathcal F|$. Unfortunately,
when varying $\kappa$ and $\nu$, either together or
independently, there is no saddle point for $\mathcal F^{(4)}_{\rm RG\Sigma}$
that would allow
one to eliminate these ambiguities by a principle of minimal
sensitivity \cite{Stevenson:1981vj}.

The renormalization group equation (\ref{dF0}) states that the
thermodynamic potential is independent of the renormalization
scale $\bar\mu$. When one uses this equation to incorporate all
the logarithmic contributions coming from higher order perturbation
theory whose form is implied by this equation, one indeed finds
the dependence of a perturbative approximation to $\mathcal F$
on $\bar\mu$ being diminished, as expected. However, we have
identified another source of ambiguity (characterized by $\kappa$
and $\nu$) which leads to large variations of the perturbative
RG$\Sigma$ result for $\mathcal F$. This highlights the numerically strong
scheme dependence of the perturbative result which still remains
after having eliminated the strong dependence on the renormalization
scale $\bar\mu$ by the RG$\Sigma$ method. (There are, of course,
other ambiguities in the RG$\Sigma$ result that could arise due
to changing the renormalization scheme from $\overline{{\rm MS}}$;
these we have not addressed.)

As can be seen from Fig.~\ref{figkappab},
the extra dependence on the parameter $\nu$ associated with
the part of the perturbative series involving half-integer
powers in $\alpha_s$ is in fact the one which dominates
the uncertainties of the perturbative result (the short-dashed lines
are rather flat when only $\kappa$ is varied, but depend strongly
on $\nu$). This part of the perturbative series is exclusively associated
with ``soft'' collective phenomena such as screening and Landau damping
and calls for a more complete treatment than conventional
perturbation theory is able to achieve. Recent attempts in this direction
that have been put forward include separate Pad\'e approximations
to soft and hard contributions \cite{Cvetic:2002ju}, optimization of
perturbation theory using the hard-thermal-loop effective action
\cite{ABS}, and approximately self-consistent propagator
resummation in the so-called $\Phi$-derivable approach 
\cite{BIR}.

\begin{acknowledgments}

Financial support has been provided by NSERC. D.~G.~C.~McK. would like
to thank the Technical University of Vienna where this work was performed.
R.~and D.~MacKenzie provided helpful suggestions.
Y.~Schr\"oder and G.~Moore pointed out typographical errors in
the first version of this paper.

\end{acknowledgments}

\appendix*
\section{}

\begin{figure}[b]
\includegraphics[bb=50 180 540 525,width=8.7cm]{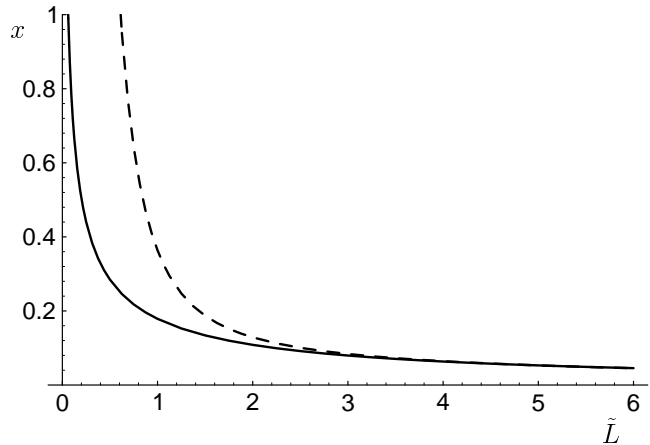}
\caption{\label{xlL} Comparison of the two-loop running coupling 
$x=\alpha_s/\pi$ resulting from the ``exact'' result (\ref{xW}) (solid line) 
with the strictly perturbative
one from Eq.~(\ref{xpt}) (dashed line)
for the case of QCD with $n_f=0$.}
\end{figure}

The beta function to two-loop order,
\be\label{twoloopbeta}
\mu^2 {dx\0d\mu^2}={dx\0dL}=b_2 x^2+b_3 x^3,
\ee
can be integrated in closed form
\be
\int_{L_0}^L dL'=L-L_0=
\int_{x_0}^x {dx'\0b_2 x'^2+b_3 x'^3}=
{1\0b_2}\left[\rho \ln{1+\rho x\0x}-{1\0x}\right]_{x_0}^x \quad (\rho=b_3/b_2)
\ee
which becomes, if $x_0\to\infty$ as $L_0\to \ln\Lambda^2$,
\be
\tilde L\equiv\ln{\mu^2\0\Lambda^2}=
{\rho\0b_2}\ln\left[We^W(-\rho e)\right]-{\rho\0b_2}\ln\rho
\ee
where
\be
W=-1-{1\0\rho x},\quad\mbox{or}\quad x=-[\rho(W+1)]^{-1}.
\ee
Hence, the solution to Eq.~(\ref{twoloopbeta}) can be written
\cite{Gardi:1998qr} as a Lambert W function \cite{Corless:1996} with
\be\label{xW}
W(z)\,e^{W(z)}=z={-1\0e}\left(\mu^2\0\Lambda^2\right)^{b_2/\rho},
\ee
where the real branch $W_{-1}$ with $W<-1$ has to be taken.

On the other hand, the standard perturbative two-loop result
is usually given as
\be\label{xpt}
x={-1\0b_2\tilde L}\left(1+{\rho\0b_2}{\ln\tilde L\0\tilde L}\right)
\ee
which assumes small $x$ and correspondingly large $L$.

In Fig.~\ref{xlL} we compare the ``exact'' two-loop result (\ref{xW})
for $x\equiv \alpha_s/\pi$ (solid line) with the strictly perturbative
one from Eq.~(\ref{xpt}) (dashed line)
for the case of QCD with $n_f=0$ ($b_2=-11/4$, $\rho=51/22$)
as a function of $\tilde L$. The divergence at $\tilde L=0$
makes itself felt significantly earlier (as $\tilde L$
approaches zero) in the standard
perturbative result (\ref{xpt}) than in the exact two-loop result (\ref{xW}).
However, for $\tilde L\gtrsim 4$, the difference
between the two coupling is less than 1.5\%. In the perturbative
treatment of hot QCD, if one chooses a renormalization scale $\bar\mu=2\pi T$
this is indeed the case for all $T>T_c\approx \Lambda_{\overline{\rm
MS}}$ (in pure glue QCD a typical value, which we have adopted
in this paper, is $T_c=1.14 \Lambda_{\overline{\rm
MS}}$); a noticeable difference thus arises only for smaller choices
of $\bar\mu/T$ in combination with $T$ sufficiently close to $T_c$. 


\end{document}